\documentclass[12pt]{article}%
\usepackage{amsmath}
\usepackage{amssymb}
\usepackage{graphicx}
\usepackage{url}
\usepackage{subfigure}
\usepackage{color}
\usepackage{geometry}
\usepackage{setspace}
\usepackage{amsfonts}%
\providecommand{\U}[1]{\protect\rule{.1in}{.1in}}
\newtheorem{theorem}{Theorem}

\newtheorem{definition}[theorem]{Definition}

\newtheorem{notation}[theorem]{Notation}

\newtheorem{remark}[theorem]{Remark}

\begin{document}

\title{Decentralization of a Machine: Some Definitions}
\author{Pradeep Dubey\thanks{Stony Brook Center for Game Theory; and Cowles Foundation
for Research in Economics, Yale University}\ \ }
\date{9 February 2015{}}
\maketitle

\begin{abstract}
We define some notions of the decentralization of a deterministic input-output
machine. This opens the possibility for introducing game-theoretic elements --
such as strategic players --- \emph{inside} the machine, as part of its
design. 


\end{abstract}

\section{Introduction}

A key feature of \textquotedblleft decentralization\textquotedblright\ is that
a complex system can be broken into smaller constituents, each of which
functions on the basis of\textit{ }information that is much more
\textit{limited }than the total information prevalent in the system. Thus
decentralization is particularly useful when information is costly to
disseminate or assimilate. A prominent example (see, \textit{e.g.,}
\cite{Barut:1998}) is that of a free market economy, where each agent simply
optimizes against prices, ignoring all his competitors; and yet an efficient
trade is achieved collectively in equilibrium. Another example (see,
\textit{e.g.,} \cite{Baye:1993}) involves \textquotedblleft local economic
networks\textquotedblright\ in which each individual interacts with a small
set of neighbors, oblivious of the rest of the participants, but the
ramifications of his actions can be felt throughout the system.

The purpose of this note is to explore the possibility of decentralization,
not in a traditional economic context, but in the design of \textquotedblleft
machines\textquotedblright. We restrict attention to a machine$\ f$ which maps
finitely many inputs to outputs. A design for $f$ consists of
smaller\ machines $\alpha$ arranged in a hierarchy. Each $\alpha$ receives as
input the outputs produced by a subset of machines lower down in the
hierarchy. Based upon its input, $\alpha$ in turn decides what output to
produce, and which machines higher up in the hierarchy to transmit the output
to. The design must, of course, implement $f$ in the following sense: for
every initial input sent into the bottom level of the hierarchy, the output at
the top level is in accordance with $f.$ Given any design, we consider the
costs of internal communication between its machines, as well as their
operating costs. An appropriate decentralization of the machine $f$ is a
design that minimizes total cost among all designs that can implement $f.$

We also point out how certain designs can have game-theoretic structures
embedded in them.

\section{The Design of a Machine}

We restrict attention to a deterministic input-output machine.

A sequence of $0^{\prime}$s and $1^{\prime}$s will be called a $01$-sequence,
for short. W.l.o.g.\footnote{See remarks \ref{finite symbols} and
\ref{Coding}} the inputs and outputs can be taken to be $01$-sequences of
length $n.$ Denoting the set of such sequences by $Seq=$ $\left\{
0,1\right\}  ^{n},$ the \textit{machine} is specified by a function
\[
f:D\longrightarrow Seq
\]
on a domain $D\subset Seq;$ which decomposes into its component functions, or
\emph{elementary machines}%
\[
f_{i}:D\longrightarrow\left\{  0,1\right\}
\]
for $1\leq i\leq n.$

Our aim is to build up the machine $f$ using \emph{smaller} elementary
machines $\alpha.$ Each such $\alpha$ receives a $0$ or a $1$ from some subset
of its predecessors, by way of input. \emph{Conditional} on its input,
$\alpha$ decides whether to produce $0$ or $1$ as output; and, furthermore, it
decides which\emph{ }of its successors to transmit the output to. Thus,
looking back from $\alpha,$ the input of $\alpha$ can be $01$-sequences of
different lengths, indexed by the subset of $\alpha$'s predecessors that
transmitted their outputs to $\alpha$.

To be more explicit, consider nonempty, disjoint, finite sets $N_{t}$ in
\textquotedblleft time periods\textquotedblright\footnote{\textquotedblleft
Time\textquotedblright\ is just a metaphor for arranging machines in layers
that are totally ordered.} $t\in\Gamma=\left\{  1,\ldots,T\right\}  $. The
initial set $N_{1}=\left\{  \eta_{1},\ldots,\eta_{n}\right\}  $ and the
terminal set $N_{T}=\left\{  \tau_{1},\ldots,\tau_{n}\right\}  $ are both of
size $n.$. For $2\leq t\leq T-1,$ the intermediate sets $N_{t}$ can be of
arbitrary size.

There is a \emph{directed graph} $G$ on the node set $N=N_{1}\cup\ldots\cup
N_{T}$. If $(\alpha,\beta)$ is a directed edge from $\alpha\in N_{l}$ to
$\beta\in N_{t}$, we require $l<t;$ i.e, the arrow of time points forward. For
any node $\alpha,$ denote its predecessor set by
\[
P_{\alpha}=\left\{  \beta:(\beta,\alpha)\in G\right\}
\]
and its successor set by%
\[
S_{\alpha}=\left\{  \beta:(\alpha,\beta)\in G\right\}
\]

Clearly $P_{\alpha}$ is empty for $\alpha\in N_{1};$ and $S_{\alpha}$ is empty
for $\alpha\in N_{T}.$ Furthermore $P_{\alpha}$ \textit{may} be empty for some
$\alpha\in N_{T}.$ For each intermediate node $\alpha\in N_{2}\cup\ldots\cup
N_{T-1}$, we require both $P_{\alpha}$ and $S_{\alpha}$ to be nonempty$.$

\begin{notation}
$N=N_{1}\cup\ldots\cup N_{T}$ and $N_{-}=N_{2}\cup\ldots\cup N_{T}$
\end{notation}

Each element $\alpha\in N_{-}$ corresponds to an elementary machine. Elements
of $N_{1}$, in contrast, are \textquotedblleft dummy\textquotedblright%
\ machines and will play the special role of \textquotedblleft
initializing\textquotedblright, or starting off, the computational process.

As was said, an input $\nu$ of $\alpha\in N_{-}$ consists of $0^{\prime}s$ and
$1^{\prime}s$ indexed by the subset $V\subset P_{\alpha}$ of $\alpha$'s
predecessors, from whence they came. Based upon $\nu$, the elementary machine
$\alpha$ first produces an output of $0$ or $1$; and then transmits this
output to a subset $W(\nu)\subset S_{\alpha}$ of its successors

\begin{notation}
$.$For any nonempty finite set $A,$ let $SeqA$ denote the set of all
$01$-sequences whose elements are indexed by $A$ (i.e., $SeqA$ is the set of
all maps from $A$ to $\left\{  0,1\right\}  );$ and when $A$ is the empty set,
$SeqA$ denotes a singleton set, consisting of the \textquotedblleft empty
sequence\textquotedblright\ $\varnothing$ which signifies the absence of any
input. Also denote $\mathcal{S}\mathfrak{(}A)=\cup\left\{  SeqB:B\subset
A\right\}  .$
\end{notation}

Given $s=(s_{1},\ldots,s_{n})\in D,$ the output of all $\alpha\in N$ is
determined recursively as follows. First, the output of $\eta_{i}\in N_{1}$ is
defined to be the component $s_{i}$ of $s\in D.$ Next, suppose the output of
every $\beta\in\cup\left\{  N_{l}:1\leq l\leq t\right\}  $ has been
determined. Consider $\alpha\in N_{t+1}.$ Define $\mathcal{D}_{\alpha}%
\subset\mathcal{S}(P_{\alpha})$ to be the set of $01$-sequences, including
possibly the empty sequence $\varnothing$, that can be received by $\alpha$
from $P_{\alpha}$ as we vary over all $s\in D.$ The output of $\alpha$ is
determined from its input in accordance with a given function, or
\emph{program,}
\[
\pi_{\alpha}:\mathcal{D}_{\alpha}\longrightarrow\left\{  0,1\right\}  ;
\]
and, for $v\in\mathcal{D}_{\alpha}$, the output $\pi_{\alpha}(v)$ is sent out
to successor machines $\varphi_{\alpha}(v)\subset S_{\alpha}$ in accordance
with a given \emph{transmission rule}%
\[
\varphi_{\alpha}:\mathcal{D}_{\alpha}\longrightarrow Pow(S_{\alpha})
\]
where $Pow(S_{\alpha})$ denotes the power set of $S_{\alpha}.$ If $\beta\in
S_{\alpha}\diagdown\varphi_{\alpha}(v),$ then it is understood that the empty
sequence $\varnothing$ has come to $\beta$ from $\alpha,$ \emph{without}
incurring any cost of transmission (i.e., without rendering the edge
$(\alpha,\beta)$ \textquotedblleft active\textquotedblright\ in the graph $G$.
In particular, if $\varphi_{\alpha}(v)$ is the empty set, then every $\beta\in
S_{\alpha}$ gets $\varnothing$ from $\alpha$ at no cost when $\alpha^{\prime
}s$ input happens to be $\nu.$

We assume that $S_{\alpha}=\cup\left\{  \varphi_{\alpha}(v):v\in
\mathcal{D}_{\alpha}\right\}  $ to rule out irrelevant elements from
$S_{\alpha}.$

\begin{definition}
$\mathcal{N}=$ $\left\{  G,\left\{  \pi_{\alpha},\varphi_{\alpha}\right\}
_{\alpha\in G}\right\}  $ is called a \emph{design}. We say that $\mathcal{N}$
\emph{implements }$f$ if
\[
f_{i}(s)=\text{ output of }\tau_{i}%
\]
for every $s=(s_{1},\ldots,s_{n})\in D$ and $\tau_{i}\in N_{T}.$
\end{definition}

\section{Informational Complexity}

\subsection{Fixed Costs}

The edges of $G$ correspond to routes for transmitting information inside the
design $\mathcal{N}=$ $\left\{  G,\left\{  \pi_{\alpha},\varphi_{\alpha
}\right\}  _{\alpha\in G}\right\}  .$ Assuming that each route costs one unit
of a \textquotedblleft red\textquotedblright\ currency to build, the (red)
\emph{fixed cost }of the design is given by
\[
c_{F}(N)=\text{\# of edges of }G
\]

Notice that, for any node $\alpha$ in $G$,%
\[
\text{\# of edges of }G\text{ leading into }\alpha=\text{cardinality of
}P_{\alpha}%
\]
is a measure of the extent of information $I_{\alpha}$ that is needed (as $s$
varies over $D)$ by $\alpha$ in order to execute its program $\pi_{\alpha}.$
By summing $I_{\alpha}$ over all the elementary machines $\alpha,$ we get
$c_{F}(\mathcal{N}).$ Thus $c_{F}(\mathcal{N})$ represents the (fixed costs)
\emph{informational complexity }of $\mathcal{N}$; and the smaller
$c_{F}(\mathcal{N})$ is, the more decentralized $\mathcal{N}$ may be thought
to be. This leads us to define
\[
\kappa_{F}(f)=\text{ min}\left\{  c_{F}(\mathcal{N)}:\text{the design
}\mathcal{N}\text{ implements }f\right\}
\]

For any given integer $l,$ it would be interesting to catalogue machines $f$
of \emph{informational complexity }$l$, along with an accompanying design
$\mathcal{N}$ that achieves $\kappa_{F}(f)=l..$

\subsection{Variable Costs}

After the fixed cost has been incurred to set up a design $\mathcal{N}=$
$\left\{  G,\left\{  \pi_{\alpha},\varphi_{\alpha}\right\}  _{\alpha\in
G}\right\}  $ to implement $f$, there is a variable cost of operating
$\mathcal{N}$. Suppose it costs one unit of a \textquotedblleft
blue\textquotedblright\ currency to transmit an \textquotedblleft information
bit\textquotedblright\ from one machine to another in $\mathcal{N}$., i.e., to
send output $1$ or $0$ from $\alpha$ to $\beta$ via the directed edge
$(\alpha,\beta)\in G.$ For ease of notation, suppose further that all inputs
$s\in D$ arrive at $\mathcal{N}$ (i.e., at the initial nodes $N_{1}$ in
$\mathcal{N)}$ with the same frequency\footnote{Otherwise, consider
$c_{V}(\mathcal{A)=}%
{\displaystyle\sum\limits_{s\in D}}
\pi_{s}\sigma_{s}(G)$ where $\pi_{s}$ is the probability of the occurence of
$s.$}.

For any $s\in D$, let $\sigma_{s}(G)$ denote the number of edges in $G$ that
are rendered active when the initial input $s\in D$ is fed into $\mathcal{N}.$
The (blue) \emph{variable} \emph{cost} of communication in $\mathcal{N}$ is
then
\[
c_{V}(\mathcal{N)=}%
{\displaystyle\sum\limits_{s\in D}}
\sigma_{s}(G)
\]

We shall refer to the pair $c_{F}(\mathcal{N}),$ $c_{V}(\mathcal{N)}$ as the
\emph{full} \emph{informational complexity }of $\mathcal{N}.$

\section{Programming Complexity}

Fix a design $\mathcal{N}=\left\{  G,\left\{  \pi_{\alpha},\varphi_{\alpha
}\right\}  _{\alpha\in G}\right\}  $ that implements $f:D\longrightarrow Seq$.
For any initial input $s\in D$ into; $\mathcal{N},$ denote by $s_{\alpha}%
\in\mathcal{D}_{\alpha}$ the input that is received by $\alpha.$ Let us think
of an \textquotedblleft algorithm\textquotedblright\ $\Gamma_{\alpha}$ for
$\pi_{\alpha}$ which inspects $s_{\alpha}\in\mathcal{D}_{\alpha}$ term by term
(according to some rule intrinsic to $\Gamma_{\alpha}$) and determines
$\pi_{\alpha}(s_{\alpha})$ in $\beta_{\Gamma_{\alpha}}(s_{\alpha})$
inspections. (If $s_{\alpha}$ is the empty sequence $\varnothing$, then
$\beta_{\Gamma_{\alpha}}(s_{\alpha})$ is taken to be $0,$ since the output
$\pi_{\alpha}(\varnothing)$ is predetermined independent of the initial input
$s,$ and so requires zero inspections of $s.$) From our point of view, the
algorithm for $\pi_{\alpha}$ is \emph{completely characterized} by the
integers $\left\{  \beta_{\Gamma_{\alpha}}(s_{\alpha}):s\in D\right\}  .$ We
may therefore assume that there is a \textit{finite }set $\mathfrak{T}%
_{\alpha}$ of algorithms for $\pi_{\alpha}$ from which to choose.in the
minimization problem displayed below.

The \emph{programming complexity} (or, \textquotedblleft green
cost\textquotedblright) of the design $\mathcal{N}$ is given by (assuming
again, for simplicity, that all $s\in D$ are equally likely)%
\[
c_{p}(\mathcal{N})=\min\left\{
{\displaystyle\sum\limits_{\alpha\in G}}
\sum_{s\in D}\beta_{\Gamma_{\alpha}}(s_{\alpha}):\Gamma_{\alpha}%
\in\mathfrak{T}_{\alpha}\right\}
\]

It is worth noting that the minimizer in the above display will choose for
$\pi_{\alpha}$ not an algorithm that does best in the worst case on
$\mathcal{D}_{\alpha},$ but rather \textit{on average, }putting more weight on
those sequences $s_{\alpha}$ in $\mathcal{D}_{\alpha}$ that occur frequently
as inputs at $\alpha$ when $s$ varies over $D.$ Also note that $c_{P}$ is in
the spirit of a \emph{variable} cost.

One might ask how to define an analogue of fixed (or, set-up) costs for
programs. We are just being speculative, but here it would seem necessary to
take into account the fact that one might be able to go from one program
$\pi_{\alpha}$ to another $\pi_{\alpha}^{\prime}$ by a modification costing
$m$ \textquotedblleft yellow\textquotedblright\ dollars; so that if
$\pi_{\alpha}$ cost $k$ such dollars to set up, then one can set up both
$\pi_{\alpha}$ and $\pi_{\alpha}^{\prime}$ for $k+m$ of those dollars. Thus
the partial order in which the programs are set up will be relevant to the
total fixed cost of setting up all the programs. But we shall ignore this
aspect of costs altogether.

\section{Decentralization of a Machine}

For non-negative numbers (weights) $x,y,z$ with $x+y+z=1,$ define the combined
complexity (cost) of $\mathcal{N}$ by
\[
c(\mathcal{N)}=xc_{F}(\mathcal{N)}+yc_{V}(\mathcal{N)}+zc_{P}(\mathcal{N)}%
\]

\begin{definition}
For any given $c,$ the \emph{complexity }of a machine $f$ is
\[
\kappa(f)=\text{ min}\left\{  c(\mathcal{N)}:\text{the design }\mathcal{N}%
\text{ implements }f\right\}  ;
\]
and the concommitant \emph{decentralization} of $f$ is a design which achieves
the minimum in the above display.
\end{definition}

To check that the definition of complexity makes sense, and that the minimum
is achieved by at least one design, it suffices to display a design that
implements $f.$ But this is obvious. Let there be just two time periods, so
that
\[
N=N_{1}\cup N_{2}=\left\{  \eta_{1},\ldots,\eta_{n}\right\}  \cup\left\{
\tau_{1},\ldots,\tau_{n}\right\}
\]

Let $G$ consist of all $n^{2}$ edges $(\eta_{i},\tau_{j})$, where $1\leq i\leq
n$ and $1\leq j\leq n$; further let $\pi_{\eta_{j}}=f_{j}$ and $\varphi
_{\eta_{j}}(v)=N_{2}.$

\section{Remarks}

\begin{remark}
\label{finite symbols} In place of $\left\{  0,1\right\}  ,$ one may consider
an arbitrary finite set $S$ of symbols. The entire foregoing (and,
forthcoming) discussion holds mutatis mutandis. The machine is now given by
$f:D\longrightarrow S^{n}$ where $D\subset S^{n};$ and $\mathcal{S}(P_{\alpha
})$ means the set of sequences in $S$ indexed by elements of subsets of
$P_{\alpha}$; and each $\pi_{\alpha}$ maps the relevant domain $\mathcal{D}%
_{\alpha}\subset\mathcal{S}(P_{\alpha})$ into $S.$
\end{remark}

\begin{remark}
\label{Coding} Any machine $M$ which maps a finite set of inputs into outputs
can be put into the format $f:D\longrightarrow Seq$ for some $D\subset Seq$ by
coding, i.e, a one-to-one map of the inputs and outputs into $01$- sequences
of suitable length $n.$ Many different functions $f$ can now represent $M$,
depending on the coding. Thus it makes sense to define $\kappa_{X}%
(M)=\min\left\{  \kappa_{X}(f):f\text{ represents }M\right\}  $; this also
defines \textquotedblleft optimal\textquotedblright\ coding.
\end{remark}

\begin{remark}
\label{geometry}It would be interesting to explore the connection between the
algebraic structure of $f$ and the geometrical structure of its decentralized design(s).
\end{remark}

\begin{remark}
\label{storage} If an output of $\alpha\in N_{l}$ is transmitted to $\beta\in
N_{t}$ for some $t>l,$ one might argue that the output has to be put in
storage for fot $t-l$ periods (at $\alpha$ or at $\beta$ or between the two).
In this case storage costs, which we have ignored, might have to be factored
into the total cost.
\end{remark}

\begin{remark}
\label{approximation} Our notion of informational complexity is based on the
worst case scenario, since we try to implement $f(s)$ for\emph{ all }$s\in D.$
We could instead consider approximate implementation, e.g., for 95\% of the
sequences in $D.$ This might reduce the complexity by a huge amount by getting
rid of the few bad elements which were very costly to handle.
\end{remark}

\begin{remark}
\label{compatibility}There is no immediate tension between informational
complexity and programming complexity. Indeed as informational complexity
reduces, the length of inputs will fall for several $\mathcal{D}_{\alpha}$,
which in turn will tend to make the programs $\pi_{\alpha}$ less complex.
Inspite of this apparent complementarity, there may be trade-offs between the
two complexities. We leave this for future inquiry.
\end{remark}

\section{Game-theoretic Design}

Suppose $\varphi_{\alpha}(v)=S_{\alpha}$ for all $v$ and consider the general
context of remark \ref{finite symbols}. Each elementary machine $\alpha$ in a
design may be regarded as a \textquotedblleft player\textquotedblright\ in a
game who is choosing a best reply to the actions of its rivals in $P_{\alpha}$
that it receives by way of input. If two elementary machines $\alpha$ and
$\beta$ are identical (possibly after a relabeling of strategies and players),
then $\alpha$ and $\beta$ may be regarded as the same player,
\textit{provided} $\pi_{\alpha}$ and $\pi_{\beta}$ are both derivable from the
same \textquotedblleft payoff function\textquotedblright. Thus introducing
players into an design $\mathcal{N}$ places considerable structure on
$\mathcal{N}.$ From the perspective of this paper, there is no game given
apriori, but only the machine $f$. We are at liberty to invent players, along
with their strategies and payoff functions --- indeed to invent the whole game
--- provided it leads to an efficacious design for implementing $f.$. It is
evident that many kinds of game-theoretic structures fit into what we have
called \textquotedblleft design.\textquotedblright\ For instance, consider a
finite normal form game. Given an $n$-tuple of (pure) strategies, let each of
the $n$ players make a best reply to the $n$-tuple. This yields a new $n$
-tuple. Iterate the procedure $k$ times. This conforms to a design. In fact we
could allow differential time lags of information, some players becoming aware
of their rivals' revisions earlier than others, or even each player observing
his different rivals with varying time lags. This ,too, is a design.

One might wonder which kinds of machines $f$ are amenable to game-theoretic
designs, but this is a topic for future exploration.

\end{document}